\documentclass[conference]{IEEEtran}
\usepackage{cite}
\usepackage{amsmath,amssymb,amsfonts}
\usepackage{algorithmic}
\usepackage{graphicx}
\usepackage{textcomp}
\usepackage{xcolor}
\def\BibTeX{{\rm B\kern-.05em{\sc i\kern-.025em b}\kern-.08em
    T\kern-.1667em\lower.7ex\hbox{E}\kern-.125emX}}
\usepackage{enumitem}
\usepackage{url} 
\usepackage[papersize={8.5in,11in}, left=0.75in, right=0.75in, top=0.80in, bottom=1.1in]{geometry}
\begin{document}

\title{Automated Vulnerability Detection in Source Code
Using Deep Representation Learning
}

\author{\IEEEauthorblockN{Christoforos Seas}
\IEEEauthorblockA{\textit{Computer Science} \\
\textit{University of Cyprus}\\
Nicosia, Cyprus \\
christoforosseas@gmail.com}
\and
\IEEEauthorblockN{Glenn Fitzpatrick}
\IEEEauthorblockA{\textit{Computer Science and Engineering} \\
\textit{Texas A\&M University}\\
College Station, TX, USA \\
glenn1fitz@tamu.edu}
\and
\IEEEauthorblockN{John A. Hamilton, Jr.}
\IEEEauthorblockA{\textit{Computer Science and Engineering} \\
\textit{Texas A\&M University}\\
College Station, TX, USA \\
hamilton@tamu.edu}
\and
\IEEEauthorblockN{Martin C. Carlisle}
\IEEEauthorblockA{\textit{Computer Science and Engineering} \\
\textit{Texas A\&M University}\\
College Station, TX, USA \\
carlislem@tamu.edu}
}

\maketitle
\begingroup
\renewcommand\thefootnote{}
\footnotetext{
© 2024 IEEE.  Personal use of this material is permitted.  Permission from IEEE must be obtained for all other uses, in any current or future media, including reprinting/republishing this material for advertising or promotional purposes, creating new collective works, for resale or redistribution to servers or lists, or reuse of any copyrighted component of this work in other work. C. Seas, G. Fitzpatrick, J. A. Hamilton and M. C. Carlisle, "Automated Vulnerability Detection in Source Code Using Deep Representation Learning," 2024 IEEE 14th Annual Computing and Communication Workshop and Conference (CCWC), Las Vegas, NV, USA, 2024, pp. 0484-0490, doi: 10.1109/CCWC60891.2024.10427574. https://ieeexplore.ieee.org/document/10427574.
}
\endgroup

\begin{abstract}
Each year, software vulnerabilities are discovered, which pose significant risks of exploitation and system compromise. We present a convolutional neural network model that can successfully identify bugs in C code. We trained our model using two complementary datasets: a machine-labeled dataset created by Draper Labs using three static analyzers 
and the NIST SATE Juliet human-labeled dataset 
designed for testing static analyzers. In contrast with the work of Russell et al. 
on these datasets, we focus on C programs, enabling us to specialize and optimize our detection techniques for this language. After removing duplicates from the dataset, we tokenize the input into 91 token categories. The category values are converted to a binary vector to save memory. Our first convolution layer is chosen so that the entire encoding of the token is presented to the filter. We use two convolution and pooling layers followed by two fully connected layers to classify programs into either a common weakness enumeration category or as ``clean.'' We obtain higher recall than prior work by Russell et al. on this dataset when requiring high precision. We also demonstrate on a custom Linux kernel dataset that we are able to find real vulnerabilities in complex code with a low false-positive rate.

\end{abstract}


\section{INTRODUCTION} \label{sec:introduction}

The ever-increasing reliance on software in modern systems and applications expose us to potential security risks arising from hidden flaws in code. These vulnerabilities, if left undetected, can be exploited by attackers, leading to system compromises and significant data breaches. Some vulnerabilities are publicly reported and addressed.  However numerous other vulnerabilities remain undiscovered within proprietary code, making it crucial to develop robust and automated detection methods.

Current methods for program analysis often rely on predefined rules, limiting their ability to identify more subtle errors made by programmers. With the vast availability of open-source repositories and the emergence of data-driven techniques, a new avenue opens for discovering vulnerability patterns more comprehensively.

In this research, we propose to address this challenge by harnessing machine learning (ML) techniques for the automated detection of vulnerabilities in C source code\footnotemark.
\footnotetext{While our work focuses on C, the techniques apply to any programming language.}By leveraging real-world code examples from open-source repositories, we aim to create a powerful and accurate vulnerability detection system capable of identifying a broader range of potential security flaws.

By employing machine learning algorithms, we will develop a sophisticated model that learns from the dataset's features to identify previously unseen vulnerability patterns. This model will enable us to overcome the limitations of rule-based detection methods and adapt dynamically to evolving threats.

The ultimate goal of this research is to contribute a state-of-the-art automated vulnerability detection tool that helps developers, security experts, and organizations proactively identify and remediate potential security weaknesses in their code, but also validate software code for any possible vulnerabilities. By doing so, we strive to enhance the overall security posture of software systems, reduce the prevalence of security breaches, and safeguard sensitive information from malicious exploits.

Throughout this paper, we will describe the development of our ML-based vulnerability detection system, present our experimental results, and showcase the practical application of our approach on real-world software packages. We anticipate that our findings will not only advance the field of vulnerability detection but also inspire further research in leveraging machine learning for cybersecurity in the broader software development community.
\color{black}
\section{RELATED WORK}

Various analysis tools exist for uncovering vulnerabilities in software. Static analyzers (e.g., Clang\cite{clang}) detect weaknesses without executing programs, while dynamic analyzers execute programs with different inputs to identify vulnerabilities. However, both approaches are limited by hand-engineered rules and lack full test coverage.

Symbolic execution [8] substitutes input data with symbolic values and explores all possible program paths Moreover symbolic execution is expensive and does not scale well for large programs. Machine learning (ML) has emerged as a promising approach, utilizing vast amounts of open-source code to learn vulnerability patterns directly from data.

In Russell et al.'s research paper \cite{draper_paper} , the researchers employed a machine-learning approach for vulnerability detection in source code. They combined neural feature representations, obtained from lexed function source code, with a random forest ensemble classifier. Feature-extraction techniques from natural language processing (NLP), including CNNs and recurrent neural networks (RNNs) were utilized for function-level source vulnerability classification. Russell et al.'s embedded code tokens into a compact representation, and their experiments revealed that an embedding size of \texttt{k = 13} performed optimally. Both convolutional and recurrent networks were employed for feature extraction, with max-pooling to generate fixed-size representations. The classification pipeline included dense layers and training involved addressing data imbalance. The dataset was split for training, validation, and testing, with hyperparameters tuned to enhance performance. Notably, the authors adopted an ensemble learning approach by using neural features as inputs to a random forest classifier, resulting in improved vulnerability detection results on a challenging dataset.

Previous works explored ML for vulnerability detection. Hovsepyan et al. \cite{prev_works} used SVM on a bag-of-words representation of tokenized Java code. Pang et al. \cite{bag_of_words} extended this with n-grams. Mou et al. \cite{n_grams} employed deep learning and abstract syntax tree embeddings for classification. Li et al. \cite{deep_learning} used RNNs to detect vulnerabilities related to library/API function calls. Harer et al. \cite{repair} trained an RNN to detect vulnerabilities in lexed code representations.

While these studies show promise, more research is needed to enhance vulnerability detection using ML techniques effectively. In this work, we explore the application of advanced ML models, leveraging larger and more diverse datasets to improve the accuracy and scalability of automated vulnerability detection.

To the best of our knowledge, prior to this research, no previous work, except for Russell et al.\cite{draper_paper}, has ventured into utilizing deep learning to directly extract features from source code within a substantial natural codebase for the purpose of detecting a diverse range of vulnerabilities. The constraints imposed by limited datasets, both in terms of size and diversity, in most prior studies, have constrained the applicability of their findings and hindered the exploitation of the full potential of deep learning. 
\section{DATA}

To train our ML model, we used deduplicated functions of SATE IV Juliet Test Suite \cite{juliet_dataset} and the C programs in the Draper VDISC dataset\cite{draper_dataset} introduced by Russell et al. \cite{draper_paper}. The SATE IV Juliet Test Suite contains synthetic code examples with vulnerabilities from 118 different Common Weakness Enumeration (CWE) \cite{cwe} classes and was originally designed to explore the performance of static and dynamic analyzers. The idea behind this is to see if we can find bugs in C code, using both human-labeled and static analysis tool data. Notably, our datasets initially contained C++ code, which we removed to concentrate solely on C code.

We also created an additional small test set containing 41 functions from Debian's commit history on GitHub, which we have made publicly available \cite{linuxvulns}. We included a buggy version from before a fix was committed and a clean version from after the fix was committed and labeled them manually from analyzing the patch. We use this test set to determine how well our machine learning method will handle complex, previously unseen code.

\subsection{Data tokenization}

To facilitate our model's training, we implemented tokenization of code by defining 91 distinct token categories. These categories can be summarized as follows:

\begin{enumerate}[label=\alph*.]
\item \texttt{KEYWORD} \
Every keyword is a different token, e.g., \texttt{int} could be 10, and \texttt{float} could be 11.
\item \texttt{SYMBOL}: Symbols, such as parentheses '(', ')' and specialized C symbols like '-$>$', '++', and '$>>$', are each assigned distinct token values for an in-depth understanding of code syntax.
\item \texttt{NUMBER}: Any numeric value is tokenized uniformly, without distinction between specific numbers.\footnotemark.
\footnotetext{In the context of our tokenization methodology, numerical entities are treated holistically, representing the entirety of a numeric value as a single token (e.g., '10' is tokenized as '10'). This approach contrasts with the methodology employed by Rusell et al. \cite{draper_paper}, wherein individual digits within a numerical sequence are tokenized separately (e.g., '10' is represented as '1' and '0'). }
\item \texttt{FUNCTION}: Certain C functions like 'strcpy' and 'memcpy' are assigned their own token values to capture their significance.
\item \texttt{IDENTIFIER}: Any other identifiers, including variable and function names fall into the 'IDENTIFIER' token category.
\end{enumerate}

This tokenization approach involves classifying each token into one of these five groups, resulting in a total of 91 distinct token types. This categorization enables the model to gain a comprehensive understanding of the code's structure and effectively process a wide range of token varieties.

\subsection{Data curation}

In our research, we encountered an issue with the Juliet dataset, where numerous almost identical functions were present, causing problems for our analysis. Consequently, we decided to discard those redundant functions. The dataset comprised 121,353 test cases written in C and C++, of which only 7,250 cases were used in our study. Most of the data consisted of \texttt{CLEAN} functions, which are functions without bugs. To ensure a balanced dataset, we introduced additional copies of buggy functions, creating an approximate 50\% representation of both buggy and clean functions. 

For our analysis, we employed a Convolutional Neural Network (\texttt{CNN}). We sought an alternative approach to representing tokens to handle the challenge of not considering numerical values explicitly due to their proximity. Many standard methods use ``one-hot'' vectors of length ``\texttt{x},'' where ``\texttt{x}'' is the highest token number. Each vector contains zeros, except for one position where a ``1'' is placed to represent the token's number. For example, the vector \texttt{[0, 0, 0, 1, ..., 0]} represents the token ``3'' because the ``1'' is in the 4\textsuperscript{th} position (the first position corresponds to the number 0). 

However, we opted not to adopt this approach due to its increased memory requirements for computation. Instead, we devised a solution that employed vectors of 8 values, representing the binary representations of token numbers. For each token, we added its binary vector representation to our dataset. For instance, the token ``3'' would be represented by the vector \texttt{[1, 1, 0, 0, 0, 0, 0, 0]}. This solution has the unfortunate side-effect of making the representations of the tokens not equidistant from each other; however, 
by employing this more memory-efficient approach, we were able to achieve our research objectives effectively. 

\subsection{Labels}

Rather than relying on specific CWE numbers, we organized our data into five distinct categories: \texttt{BUFFER, LOGIC, MEMORY, NUMERICAL, CLEAN}. Each of these categories corresponds to a specific type of bug. For instance, if our model identifies a bug as a \texttt{BUFFER} type, the precise CWE number, whether it's CWE120 or CWE121, becomes less relevant. What matters most is recognizing the bug as a buffer overflow issue, which aligns with our primary focus. This categorization approach not only simplifies bug classification but also aids in the training of our ML model. In Table~\ref{table:cwe_labels}, the reader can find the labels assigned to each CWE present in Russell et al.'s dataset. Furthermore, we applied the same categorization method to label the Juliet dataset based on its functional characteristics. 

\subsection{Padding}
In order to facilitate the effective utilization ofscale datasets, thereby improving the model's training efficiency.

``Function tokens'' and ``Target.'' Notably, the ``Function tokens'' column represents the sequence of tokens in each function. However, as functions may vary in the number of tokens they contain, we initially padded all functions to match the length of the function with the highest number of tokens.

Upon experimentation, we encountered challenges with the excessive data size, hindering the model's training process. To address this issue, we devised an approach where each function was limited to a maximum of 500 tokens, effectively capping the width of the ``Function tokens'' column at 500.

This preprocessing technique allowed us to maintain the essential structure of the data while mitigating the computational complexity associated with large-scale datasets, thereby improving the model's training efficiency.

\begin{table*}[htbp]
\caption{Labels used for each CWE from Russell et al.'s dataset}
\begin{center}
\begin{tabular}{|c|c|c|}
\hline
CWE ID & CWE Description & Label \\
\hline
120/121/122 & Buffer Overflow & BUFFER \\
119 & Improper Restriction of Operations within the Bounds of a Memory Buffer & BUFFER \\
476 & NULL Pointer Dereference & MEMORY \\
469 & Use of Pointer Subtraction to Determine Size &  MEMORY  \\
20, 457, etc. & Improper Input Validation, Use of Uninitialized Variable, Buffer Access with Incorrect Length Value, etc. & NUMERICAL  \\
\hline
\end{tabular}
\end{center}
\label{table:cwe_labels}
\end{table*}

\section{METHODS}



\subsection{Convolutional Neural Network Architecture}
\begin{figure*}[ht] 
	\centering
	\includegraphics[width=\textwidth]{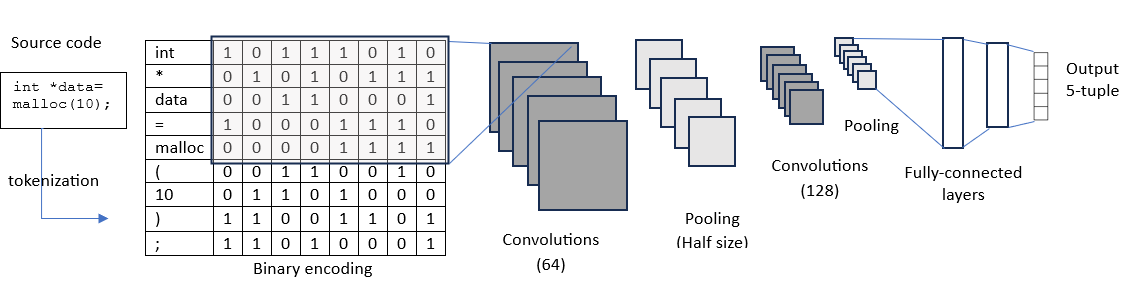}
	\caption{Convolutional Neural Representation-Learning Approach for Source Code Classification. }
	\label{fig:cnn}
\end{figure*}

As previously indicated, our model utilizes a Convolutional Neural Network (CNN) architecture for its core, as depicted in Fig.~\ref{fig:cnn}. The input is tokenized into vectors of size 8, which undergo a convolution operation using an 8x5 kernel matrix that slides vertically. The resulting output is then passed through the pooling layer for downsampling. This effectively halves the size of its input. Subsequently, the pooled data is fed into another convolutional layer, then passed again through another pooling layer, and lastly processed by two fully connected layers for classification. The first fully connected layer outputs 64 units/features to the second, and the second outputs 5 units/features representing our five distinct outputs (\texttt{BUFFER, LOGIC, MEMORY, NUMERICAL, CLEAN}). This architecture enables effective learning of meaningful representations from tokenized input, facilitating accurate and robust source code classification for various tasks. 

Researchers are encouraged to explore alternative architectural configurations, including variations in the number of channels between these layers to fine-tune their models for optimal performance in their specific tasks.

\subsection{Kernel Matrix Configuration}
The kernel matrix we have selected is sized at 8x5, effectively functioning as a 5-gram processor, allowing it to handle five tokens simultaneously. This design choice is instrumental in capturing local patterns and dependencies within the input data effectively. For a visual representation of this approach, please refer to Fig.~\ref{fig:cnn}. It's noteworthy that our kernel size choice
differs from that of Russell et al.\cite{draper_paper}, which utilized a kernel size
of $(m \times k)$ with m = 9. In our work, we opted for a kernel size of 8x5, guided by empirical evidence and task-specific requirements.



\subsection{Comparative Analysis with Prior Work}
Our data preprocessing pipeline involved extensive cleaning and curation to ensure the quality and relevance of the data, aligning it closely with the specific focus on C code vulnerability detection. Additionally, it is noteworthy that unlike Russell et al.'s approach, we did not employ a random forest classifier in our vulnerability detection system, opting instead for a different classification strategy.
Furthermore, in contrast to Russell et al.\cite{draper_paper}, which employed a total of \(n = 512\) filters in their convolutional layers, our implementation follows a different configuration. Our choice of configuration was driven by experimentation, where we found that these settings consistently produced the best results. We would encourage fellow researchers to explore and experiment with different configurations to advance the field of vulnerability detection further.

Specifically, we have utilized:
\begin{itemize}
    \item 64 filters for our first convolutional layer.
    \item 128 filters for our second convolutional layer.
    \item 64 units/features for our first fully connected layer.
    \item 5 units/features for our last fully connected layer, representing our five distinct outputs (\texttt{BUFFER, LOGIC, MEMORY, NUMERICAL, CLEAN}).
\end{itemize}


\section{RESULTS}
\begin{figure}[ht] 
	\centering
    \includegraphics[width=0.5\textwidth]{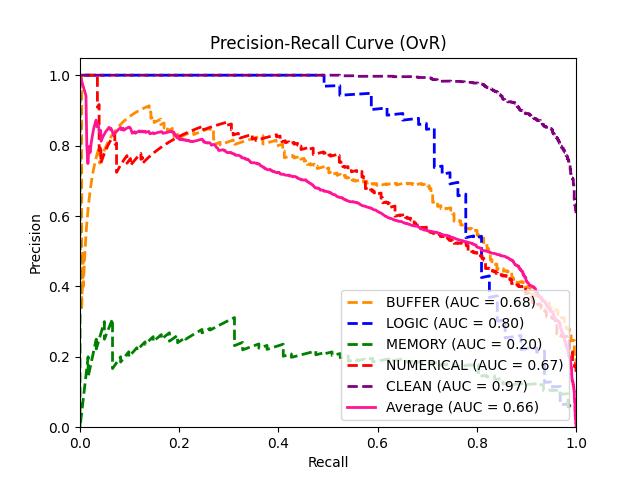}
	\caption{Multi-label PR Curve tested on Juliet dataset with our approach. Our labels are multiple CWEs grouped together, as shown in the table~\ref{table:cwe_labels}. The \texttt{CLEAN} curve is the PR curve for the functions that are not buggy.}
	\label{fig:juliet_only}
\end{figure}

After conducting training with our CNN, we proceeded to evaluate its performance on distinct datasets. Specifically, we subjected it to testing using the Juliet dataset, as depicted in Fig.~\ref{fig:juliet_only}. 

\begin{figure}[ht] 
	\centering
    \includegraphics[width=0.5\textwidth]{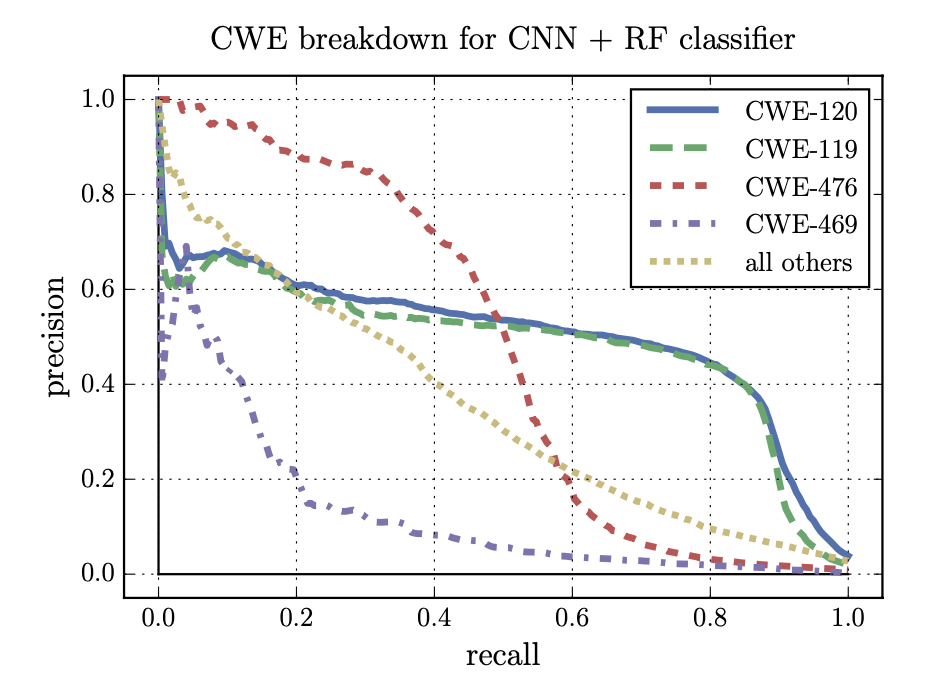}
	\caption{Multi-label PR Curve from Russell et al.'s paper \cite{draper_paper} on different CWEs}
	\label{fig:previous_graph}
\end{figure}

One of the key findings of our research is the notable enhancement in precision (true positive rate) at lower recall (total number of buggy programs analyzed) levels for specific bug categories For instance, when compared to Russell et al.’s
methodology, our approach demonstrated higher precision in
detecting some bugs at the same recall rates. To illustrate, for
recall = 0.4, our \texttt{BUFFER} precision is approximately 0.8, whereas
in Russell et al.\cite{draper_paper}, both CWE119 and CWE120 precision for
recall = 0.4 is lower than 0.6. This improvement signifies our algorithm's effectiveness in identifying these critical bug types with fewer false positives. In Fig.~\ref{fig:previous_graph}, the results from Russell et al.'s work  \cite{draper_paper} on Common Weakness Enumerations (CWEs) are presented. 

We decided to use labels to group CWE categories strategically. In particular, we labeled CWE 119-122 as 'BUFFER' to ensure uniform treatment. This approach enabled us to achieve higher precision. By categorizing these CWEs in this manner, we improved our model's ability to accurately identify and address vulnerabilities associated with specific bug types, such as buffer overflows. For the corresponding labels assigned to each CWE, please refer to Table~\ref{table:cwe_labels}. In our assessment, we focused on the Precision-Recall (PR) curve because we wanted to highlight the trade-off between missing vulnerabilities and overwhelming analysts with a high false-positive rate.

In addition to the comprehensive evaluation conducted on various datasets, we sought to demonstrate the practical effectiveness of our model in real-world scenarios. By subjecting our model to testing using historical Linux kernel bugs, we achieved a noteworthy outcome. Our model successfully identified some specific bugs as buffer overflow bugs, aligning with the known characteristics of the issues. Remarkably, the model's performance exhibited a significant disparity. When trained with 80\% of the functions from the Juliet dataset (reserving 20\% for testing), it successfully detected only one out of the 41 functions, with a confidence level of 0.7. However, when trained with the complete Juliet dataset, the model exhibited enhanced performance, identifying four functions with the same confidence level. More precisely, it recognized these functions:
\begin{itemize}
    \item \texttt{static int iscsi\_add\_notunderstood \_response (char *key, char *value, struct iscsi\_param\_list *param\_list)} This function was also detected when the model was trained using only 80\% of the Juliet dataset.  
    
    \item \texttt{static int} \\ \texttt{ttusbdecfe\_dvbs\_diseqc\_send\_master
    \_cmd} \texttt{(struct dvb\_frontend* fe, struct dvb\_diseqc\_master\_cmd *cmd)} 
    
    \item \texttt{static s32 brcmf\_cfg80211\_start\_ap (struct wiphy *wiphy, struct net\_device *ndev, struct cfg80211\_ap\_settings *settings)} 
    
    \item \texttt{static s32 \\ brcmf\_cfg80211\_escan\_handler (struct brcmf\_if *ifp, const struct brcmf\_event\_msg *e, void *data)} 
\end{itemize}

The function \texttt{ iscsi\_add\_notunderstood \_response} contained a buffer overflow on a heap-allocated object because the \texttt{strncpy} routine was called with the number of bytes in the input string, not the destination buffer size. Here is the unpatched code:

\begin{verbatim}
strncpy(extra_response->key, 
   key, 
   strlen(key) + 1);
   
strncpy(extra_response->value, 
   NOTUNDERSTOOD,
   strlen(NOTUNDERSTOOD) + 1);
\end{verbatim}

The code after the patch is:
\begin{verbatim}
strlcpy(extra_response->key, 
   key,
   sizeof(extra_response->key));
   
strlcpy(extra_response->value, 
   NOTUNDERSTOOD,
   sizeof(extra_response->value));
\end{verbatim}

\texttt{ttusbdecfe\_dvbs\_diseqc\_send\_master\_cmd} failed to check the incoming message size, creating an overflow of a stack-based buffer. The patched version of the code added the following check:

\begin{verbatim}
if (cmd->msg_len > sizeof(b) - 4)
   return -EINVAL;
\end{verbatim}

\texttt{static s32 brcmf\_cfg80211\_start\_ap} and \texttt{static s32 brcmf\_cfg80211\_escan\_handler} also failed to perform proper validation of data sizes when interacting with stack-based buffers.



This result not only underscores the robustness of our bug detection approach but also signifies its potential applicability in practical software development contexts. The model's ability to accurately detect a proven Linux kernel bug reinforces its capacity to recognize intricate vulnerabilities that may have broader implications within the software landscape.

\subsection{Practical Implications and Significance}

Our research presents an advancement to the field of software security, offering a powerful and versatile tool for software developers and security analysts alike. The most significant contribution of our model is its ability to achieve a higher precision than prior work at the same recall level. Therefore we can choose a threshold where our model excels in accurately identifying vulnerabilities when it makes a prediction, which is a pivotal advantage in real-world security applications.

In practical terms, this capability has several  implications:

\begin{enumerate}
    \item Enhanced Security Posture: Our model's high precision ensures that it can efficiently identify critical vulnerabilities without inundating users with false alarms. This characteristic is paramount in security-critical domains, where false positives can lead to unnecessary disruptions and distractions. By significantly reducing false positives, our model allows security teams to focus their attention on genuine threats, thus bolstering the overall security posture of software systems.
    
    \item Rapid Vulnerability Detection: The ability of our model to swiftly and accurately pinpoint vulnerabilities will contribute to the advancement of software development and security. Traditional methods often involve manual code reviews or rule-based systems, which are time-consuming and prone to human error. Our model automates and accelerates the vulnerability detection process, enabling developers to identify and address security issues in real-time or during the development phase, ultimately saving time and resources.
    
    \item Improved Software Reliability: Software reliability and security go hand in hand. By effectively identifying vulnerabilities early in the development cycle or during routine code reviews, our model contributes to the overall reliability of software systems. Our model reduces the risk of security breaches and enhances user trust and satisfaction.
    
    \item Scalable and Adaptable: One of the notable aspects of our model is its adaptability to different programming languages and projects. Whether it is a large-scale software application or a small project, our tool can be applied to detect vulnerabilities, providing consistent and reliable results. This adaptability makes it a versatile choice for organizations with diverse software portfolios.
    
    \item Long-Term Impact: As the software landscape evolves, so do the threats. Our research lays the foundation for ongoing advancements in automated vulnerability detection. By continuously training and refining the model with new data, it can adapt to emerging vulnerabilities and evolving coding practices, ensuring its long-term relevance and effectiveness.
\end{enumerate}

Our training process, which took a total of 9.5 hours, was conducted exclusively on a system equipped with an Intel(R) Xeon(R) CPU E7-4850 v2 running at 2.30GHz with 755 GB of RAM and without the utilization of a GPU.

\section{CONCLUSIONS}

By training a machine learning model to identify bugs in C code using both human-labeled and dataset derived from static analyzer tools, we have successfully developed a model of significant utility. This model exhibits the potential to effectively identify bugs in critical domains, such as aviation and space technologies, as well as within complex systems like operating systems. This is evidenced by its capacity to detect flaws in the Linux kernel.

Our research represents a step forward in the field of software security. Its ability to balance high precision with real-world practicality makes it an useful asset for software developers, security analysts, and organizations striving to protect their software assets and user data. By integrating our model into their workflows, practitioners can fortify their defenses, reduce security risks, and ultimately build more secure and reliable software systems, regardless of the programming language used.

Our evaluation strategy is founded on our core objective of enhancing the accuracy of detecting true positive instances. We place substantial emphasis on achieving a high true positive rate, which implies that when our model predicts a bug, we have a high degree of confidence in its accuracy. This focus stems from our priority to minimize false positives and maximize true positives, even if this leads to a lower overall recall rate. We intend to prioritize precision over recall, ensuring that the bugs detected by our model are indeed valid.

By adopting this strategy, we aim to produce a bug-detection system that provides reliable and trustworthy results. This approach aligns with our overarching goal of achieving dependable outcomes in bug identification, particularly in contexts where the certainty of detection holds paramount significance


While our initial objective centered on the detection of vulnerabilities within the Linux kernel, we have demonstrated the model's capability to identify existing vulnerabilities in this domain. This accomplishment underscores the practical utility of our model in identifying known vulnerabilities that could have far-reaching impacts if left unaddressed. While we did not discover new Linux kernel vulnerabilities in this particular instance, our research paves the way for such discoveries in the future. This advance is poised to save valuable time for software developers by enabling them to concentrate their efforts on the portions of code with a higher likelihood of containing issues.

The model's proficiency in detecting real-world bugs represents a crucial step toward enhancing software quality and reliability across various industries. As we look ahead, we envision the model serving as a strategic asset in ensuring the integrity of critical software systems, contributing to safer and more efficient technological landscapes. One place it might be used is in combination with large language models. Some exploration has been done on using large language models (LLMs) to find vulnerabilities (e.g. by Das Purba et al.\cite{svd_with_llm}). Rather than using the LLM to identify the vulnerability, we might instead use the LLM to rewrite the code after our technique has identified a vulnerable function. If this process is automated, we may be able to find and fix more vulnerabilties because the cost of false positives will be eliminated (assuming the LLM doesn't introduce a bug when ``fixing'' a function identified as vulnerable that has no vulnerabilities).

\bibliographystyle{bib_sophia}
\bibliography{references}

\end{document}